# Chaotic Quantization and the Mass Spectrum of Fermions

**Muhammad Maher*** **and Christian Beck♣**
School of Mathematical Sciences
Queen Mary, University of London
Mile End Road, London E1 4NS, UK

## Abstract

In order to understand the parameters of the standard model of electroweak and strong interactions, one needs to embed the standard model into some larger theory that accounts for the observed values. This means some additional sector is needed that fixes and stabilizes the values of the fundamental constants of nature. We describe how such a sector can be constructed using the so-called chaotic quantization method applied to a system of coupled map lattices. We restrict ourselves in this short note on verifying how our model correctly yields the numerical values of Yukawa and gravitational coupling constants of a collection of heavy and light fermions using a simple principle, the local minimization of vacuum energy.

## Introduction

String and M-theory predict an enormous amount of possible vacua after compactification. In each of these vacua the cosmological constant as well as the fundamental constants of nature have different values. This is the so-called 'landscape' picture. To select the right vacuum, one always can implement the anthropic point of view. But is an anthropic argument really the only solution?

A natural idea to help out of this dilemma would be that their should be some general principle provided by some form of an additional sector as yet not included in neither the ordinary standard model, nor in ordinary string theories. This general principle should select, fix and stabilize the standard model parameters in the way we do observe them. Such a principle may be based on chaotic dynamics.

The important role of chaos and fractals in quantum field theories and string theories has been emphasized in various recent papers and books, see for example [1-15]. Many approaches have been suggested, most notably the works of M. S. El Naschie using the so-called E-∞ theory of fractal space-time [12]. Our approach here is based on coupled map lattices [21] simulating vacuum fluctuations, which seem to give us a hint at how this additional sector could look like [8, 9]. In practice standard model parameters can be thought of as being fixed by so-called moduli fields [16]. If one knows the correct moduli potentials describing the world around us, then one knows the correct standard model parameters. So what could be a theory to construct these moduli

* Electronic address: m.maher@qmul.ac.uk

♣ Electronic address: c.beck@qmul.ac.uk

potentials? In principle the potentials should follow from the embedding theory but little is known in practice due to the enormous complexity inherent in the current mainstream theories. But as with any unknown theory one can be guided by first trying to find an empirical model that does the correct thing, i.e. reproduces the observed value of the fine structure constant and other fundamental constants, and then later try to embed it into a larger context. The interesting thing is that such an empirical model is possible [8, 9, 17]. There is a class of highly nonlinear chaotic dynamical systems that seem to reproduce the 'correct' standard model parameters by a simple selection mechanism — the minimization of vacuum energy.

Physically, the above-mentioned coupled chaotic dynamics can be regarded as describing rapidly fluctuating scalar fields associated with vacuum fluctuations. These are vacuum fluctuations different from those of QED or QCD. The most natural embedding is to associate the above chaotic vacuum fluctuations with the currently observed dark energy in the universe [17]. The chaotic fields, living in the dark energy sector, generate effective potentials for moduli fields — just the same moduli fields that are responsible for the fundamental constants of the standard model of electroweak and strong interactions. The moduli fields then move to the minima of the potentials generated by the chaotic fields, and fix the fundamental constants of nature [16]. This is why we are interested in local minima of the vacuum energy in this paper. The chaotic sector appears to provide a possible answer to the question why we do observe certain numerical values of standard model parameters in nature. It can be used to avoid anthropic considerations for fundamental constants. Moreover, it generates a small cosmological constant in a rather natural way [17].

## Chaotic Quantization

In the chaotic quantization approach one replaces the Gaussian white noise of the Parisi-Wu approach of stochastic quantization [18] by a deterministic chaotic process on a very small scale. A simple model is to generate the noise by Tchebyscheff maps $\mathbf{T_N}$. In nonlinear dynamics, $\mathbf{T_N}$ are standard examples of chaotic maps, just as the harmonic oscillator is a standard example in linear dynamics. One has $\mathbf{T}_2(\varphi) = 2\varphi^2 - 1$ for the lowest order N=2. Most important for our purposes is the property that the $\mathbf{T_N}$ have least higher-order correlations among all smooth systems conjugated to a Bernoulli shift, and are in that sense closest to Gaussian white noise, as close as possible for a smooth deterministic system [19, 20]. Any other map has more higher-order correlations. What does this mean for chaotic quantization? Actually the $\mathbf{T_N}$ are automatically selected as most ideal candidates if we argue that Gaussian white noise is chosen to be generated in nature by something deterministic chaotic on the smallest quantization scales, which aims at making the small-scale deviations from ordinary quantum mechanics as small as possible.

## Coupled Chaotic Map Lattices

Now that we assumed that the noise fields used for quantization are dynamical in origin, it'd be plausible to allow for some coupling between neighboured noise fields. Physically it is most reasonable that the coupling has a Laplacian form, since this is the most relevant coupling form in quantum field and string theories. This leads to coupled map lattices of the nearest-neighbour coupling form. We end up in the 1-dimensional case with coupled Tchebyscheff maps of order N of the form

$$\varphi^{\mathbf{i}}_{\mathbf{n}+1} = \mathbf{(1\text{-}a)}\ \mathbf{T}_N(\varphi^{\mathbf{i}}_{\mathbf{n}}) + s\frac{\mathbf{a}}{2}(\mathbf{T}^{\mathbf{b}}_N(\varphi^{\mathbf{i}+1}_{\mathbf{n}}) + \mathbf{T}^{\mathbf{b}}_N(\varphi^{\mathbf{i}-1}_{\mathbf{n}})), \qquad (1)$$

where $i$ is a 1-dimensional lattice coordinate, a $\in$ [0, 1] is a coupling constant, s = $\pm 1$, and b takes on values 0 or 1 ($\mathbf{T}^0(\varphi) = \varphi, \mathbf{T}^1(\varphi) = \mathbf{T}(\varphi)$). Our choices throughout all the calculations leading to the presented results below were s=1 (diffusive coupling) and b=0 (backward coupling). The above chaotic dynamics is deterministic chaotic spatially extended, and strongly nonlinear. The field variable $\varphi^{\mathbf{i}}_{\mathbf{n}}$ is physically interpreted in terms of rapidly fluctuating virtual momenta in units of some arbitrary maximum momentum scale [9].

## Some Numerical Calculations

### 1-Dimensional Case

We used equation (1) to calculate the self-energy $\mathbf{V} \equiv \langle \mathbf{V}(\varphi) \rangle_{\mathbf{a}}$ of the above introduced spatially extended chaotic coupled map lattice in one space-time dimension. We implemented random initial conditions $\varphi^{\mathbf{i}}_0 = \mathbf{cos}\,\pi\mathbf{u}$ , $\mathbf{u} \in [0,1]$ and also periodic boundary conditions that can be expressed as $\varphi^1_{\mathbf{n}} = \varphi^{\mathbf{I}}_{\mathbf{n}}$ where I labels the last site on the linear lattice. In the N=2 case, $\mathbf{V}(\varphi)$ can be written as $\mathbf{V}_2(\varphi) = \varphi - 2/3\varphi^3$, where we imposed that $\mathbf{V}_{\pm\mathbf{N}}(\varphi) = \mp\int \mathbf{T}(\varphi)\mathbf{d}\varphi$, which is an essential choice to facilitate the calculation of self-energy [8, 9]. Assuming ergodicity, expectations $\langle \mathbf{V}(\varphi) \rangle_{\mathbf{a}}$ for a given coupling parameter (a) are numerically calculable as time averages

$$\langle \mathbf{V}(\varphi) \rangle_{\mathbf{a}} = \lim_{\mathbf{M}\to\infty,\ \mathbf{I}\to\infty} \frac{1}{\mathbf{M\,I}} \sum_{\mathbf{n}=1}^{\mathbf{M}} \sum_{\mathbf{i}=1}^{\mathbf{I}} \mathbf{V}(\varphi^{\mathbf{i}}_{\mathbf{n}})\,. \qquad (2)$$

For a → 0 one numerically observes the scaling behaviour $\langle \mathbf{V}(\varphi) \rangle_{\mathbf{a}} - \langle \mathbf{V}(\varphi) \rangle_0 = \sqrt{\mathbf{a}}\,\mathbf{F}^{(\mathbf{N})}(\mathbf{log(a)})$ where $F^{(N)}$ is a periodic function of $\mathbf{log(a)}$ with period $\mathbf{log\,N}^2$ and $\langle \mathbf{V}(\varphi) \rangle_0 = 0$ for N=2 [9]. Such a log periodic scaling is manifested in our results shown in Fig (1) below and was analytically proved in [25]. From the periodicity it follows that if there is some local minimum of the potential at $b_i$ then there is also a minimum at $\mathbf{b_i} / \mathbf{N}^{2L}$, where L is an integer. In other words, minima are only determined modulo 4 for N=2. In one period shown in fig (2), and if we magnify enough (calculate for sufficiently dense values of the coupling a) and use sufficiently high values of the number of lattice sites and the number of iterations, one remarkably observes 9 different minima that coincide with Yukawa and gravitational couplings of 9 standard model fermions modulo 4. One observes that the minima with i = 2, 6, 10 turn out to coincide with Yukawa couplings modulo 4 of the heavy fermions τ, b, c according to the formula $\mathbf{b_i} = 1/4\,\alpha_2(\mathbf{m_H} + 2\mathbf{m_f}).(\mathbf{m_f}/\mathbf{m_w})^2 .\, 4^{\mathbf{L_f}}$ where f = τ, b, c, respectively, $\mathbf{m_H}$ is the Higgs mass taken to be 154 GeV [9], $\mathbf{m_w}$ is the mass of the W-boson, $\alpha_2(\mathbf{E})$ is the running weak coupling constant [22], and $\mathbf{L_f}$ is a suitable integer for a particular fermion f. The results are almost independent of the Higgs mass [9]. For the light fermions one observes that the self-energy has local minima for couplings that coincide with gravitational couplings modulo 4. We numerically observe for i = 4, 7, 8, 9 that $\mathbf{b_i} = 1/4(\mathbf{m_f}/\mathbf{m_{Pl}})^2 .4^{\mathbf{L_f}} .2$ where f = e, d, u, s, respectively. Here $\mathbf{m_{Pl}}$ denotes the Planck mass. Solving for $m_f$, one can thus get fermion mass predictions modulo 2. The relevant power of 2 (the value of $\mathbf{L_f}$ in the above equations of $\mathbf{b_i}$) can then be obtained from other minima and additional symmetry considerations [9]. We also observe the minima with i = 3, 11 which may yield neutrino mass predictions. As for i = 5, it is assumed to account for Yukawa interaction of the top quark while i = 1 is expected to host a doublet to account for the heaviest neutrino and the muon [8, 9]. A list of the predicted fermion masses corresponding to the confirmed local minima can be found in sections 8.6 and 8.7 in [9] and the excellent agreement (3-4 digits) with the experimentally observed masses is remarkable.

Actually it was argued in [8, 9] that there are 11 minima. We have now repeated these calculations with much higher computer power which allowed us to do highly precise calculations (see typical samples in figs 3, 4) around the values of the fermion couplings assumed in [8, 9]. Our calculations returned a firm confirmation of only 9 of them. Literally all the minima suggested in [8, 9] were confirmed except for $b_1$ and $b_5$. At $b_1$ and $b_5$ we got something more like an inflexion point (an unstable stationary point) rather than a typical stable minimum. Further calculations have been performed for different sets of initial-conditions as well as at minima valleys of lower couplings (lower $\mathbf{L_f}$) in an attempt to eliminate the ambiguities for light fermions. So far the existence of the minima $b_1$ and $b_5$ is unconfirmed. Finally we should say that the probability of such a very precise matching between our calculations and the experimental data for only two fermion masses (e.g. ***e*** and ***τ***) to be a result of mere random coincidence is of the order of one in a million.

## D-Dimensional Case

The coupled map lattices (CMLs) described in (1) can be studied on lattices of arbitrary dimension. We generalized the investigations and the numerical calculations of the 1-dimensional case to higher space-time dimensions ($4 \geq \mathbf{D} > 1$). If we take the D=4 case as an example, the CMLs can be written as:

$$\varphi_{n+1}^{h,i,j,k} = (1-a)T(\varphi_n^{h,i,j,k}) + \frac{a}{8}(\varphi_n^{h,i+1,j,k} + \varphi_n^{h,i-1,j,k} + \varphi_n^{h,i,j-1,k} + \varphi_n^{h,i,j+1,k}$$
$$+ \varphi_n^{h,i,j,k-1} + \varphi_n^{h,i,j,k+1} + \varphi_n^{h+1,i,j,k} + \varphi_n^{h-1,i,j,k}) \qquad (3)$$

Like in the D=1 case, we implemented periodic boundary conditions given by: $\varphi_n^{1,i,j,k} = \varphi_n^{H,i,j,k}$ , $\varphi_n^{h,1,j,k,} = \varphi_n^{h,I,j,k}$ , $\varphi_n^{h,i,1,k} = \varphi_n^{h,i,J,K}$ **and** $\varphi_n^{h,i,j,1} = \varphi_n^{h,i,j,K}$ where H, I, J and K are the last space-time sites along each relevant space-time direction on the cubic lattice. In this case, $T_N(\varphi), V_N(\varphi), \varphi_0$ have exactly the same form as in the 1D case and the self energy $\langle V(\varphi) \rangle_a$ is given by

$$\langle V(\varphi) \rangle_a = \lim_{M\to\infty, H\to\infty, I\to\infty, J\to\infty, K\to\infty} \frac{1}{MHIJK} \sum_{n=1}^{M} \sum_{j=1}^{J} \sum_{k=1}^{K} \sum_{i=1}^{I} \sum_{i=1}^{I} V(\varphi_n^{h,i,j,k}) \qquad (4)$$

We observed a log-periodic behaviour similar to, however apparently smoother than, the D = 1 case. This is demonstrated in fig (5). We magnified one period of the log-periodic curve in the small coupling scaling region, which is identical to scaling region considered in the D=1 case. We adopted an optimal large-enough combination of the number of lattice sites and the number of iterations to ensure reasonable statistics. The gross calculations of all the ($4 \geq \mathbf{D} > 1$) three cases took about 8000 hours of parallel computing time on the e-science computing cluster at QMUL which is part of GridPP; the UK's grid computing facility. To our surprise, rather than obtaining some structure, we got a single smooth structure-free broad minimum for all $4 \geq \mathbf{D} > 1$ cases, see fig (6) for the D = 4 case as an example. We couldn't relate the corresponding coupling of about 0.000358(4) to any of the known particle masses using similar arguments to that used in the D = 1 case. For the time being, the only hypothesis we have for the observed smoothness in the D >1 cases as compared to the wriggling fractal shape in the D =1 case (fig (2)) is that this might be an effect of what we can call an extended version of Nash embedding [23, 24]. If correct, this is rather a topological-geometrical argument to be complemented in future with a concrete physical explanation.

## Conclusion

We have studied a chaotic scalar field that lives in one space-time dimension as a model of vacuum fluctuations. Expectations of the self energy can be used to generate potentials that can fix the fundamental constants of nature. The generalization of these fields to higher space-time dimensions didn't yield any interesting structure in the shape of self energy. The nonlinear dynamics given by eq. (1) appears to distinguish certain numerical values of coupling constants that do coincide with known standard model coupling constants with very high precision [8, 9]. A random coincidence can really be excluded. In this way the chaotic fields can help to select the 'correct' vacuum out of an enormous number of possibilities to shape the world around us. Our numerics shows that anthropic arguments for standard model parameters are most likely to be wrong. This further emphasizes the physical importance of the chaotic quantization method and more importantly the deep significance of the underlying non-linear dynamics given by eq.(1).

## Figures:

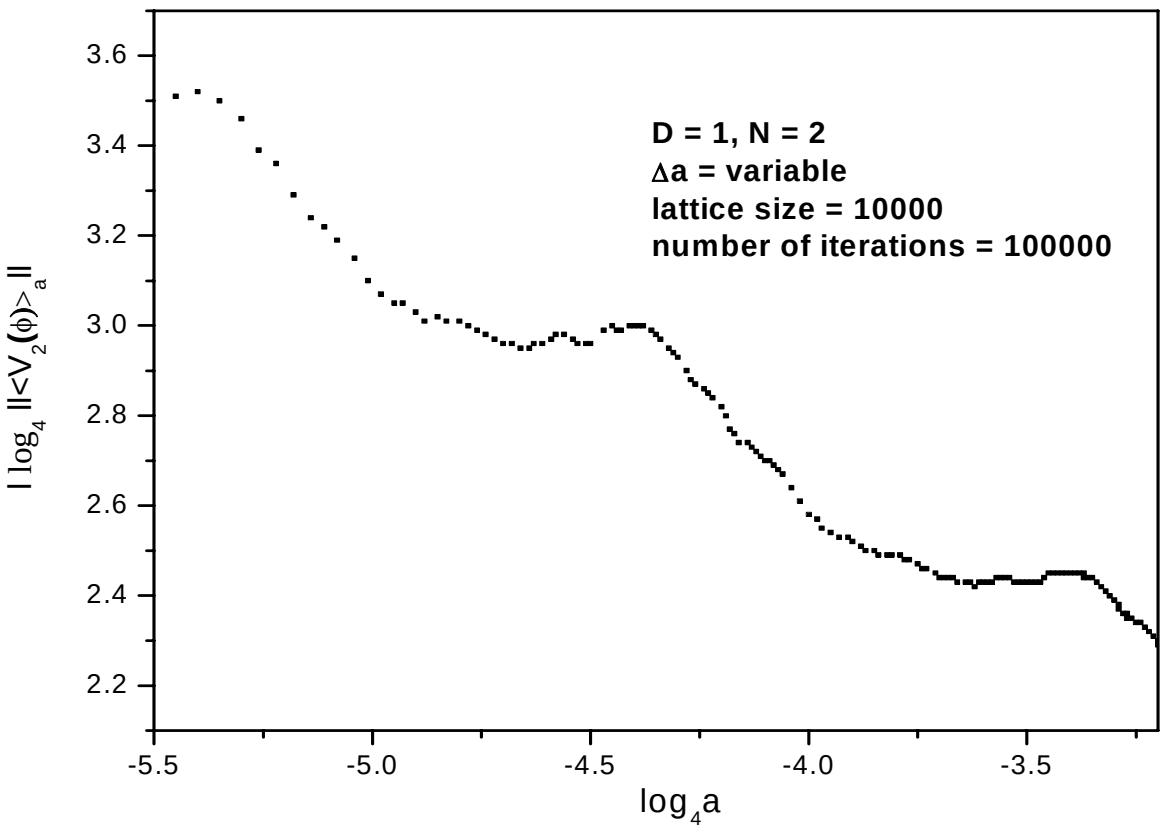

Fig (1)

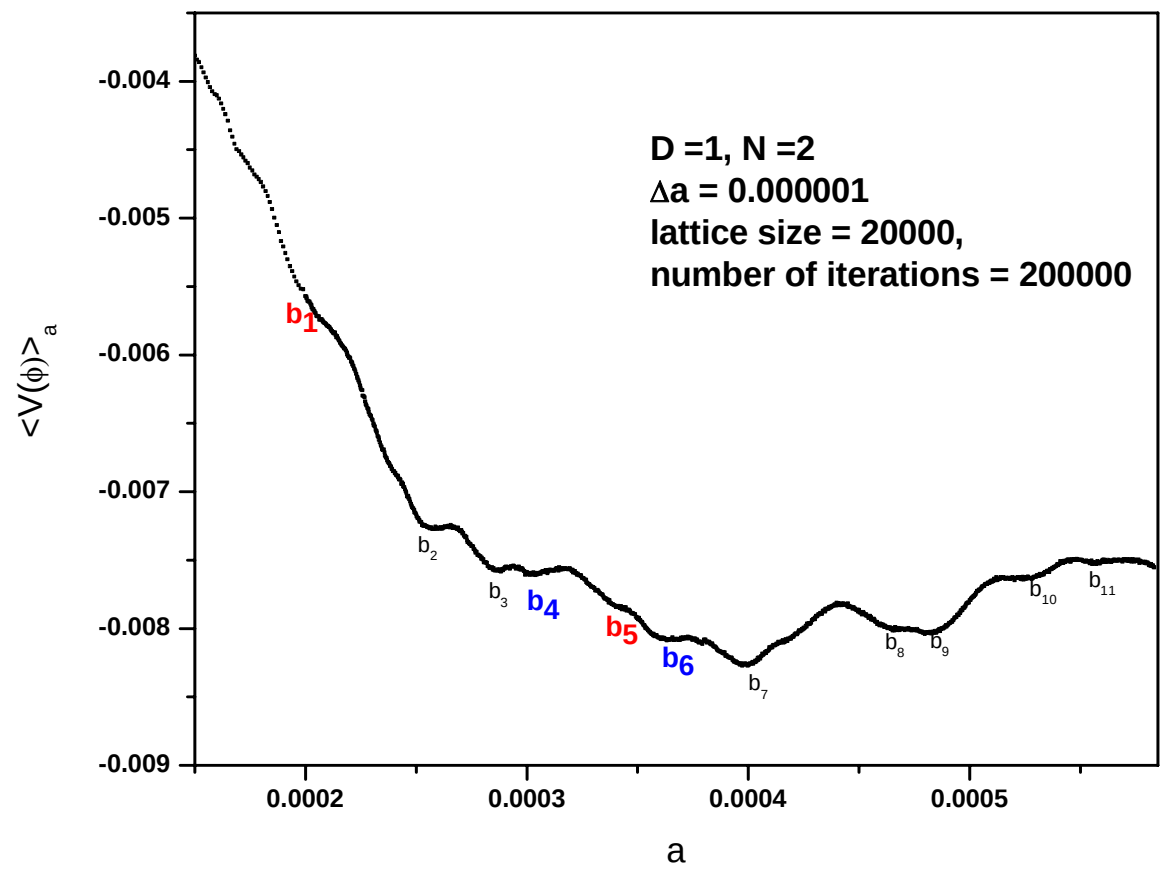

Fig (2)

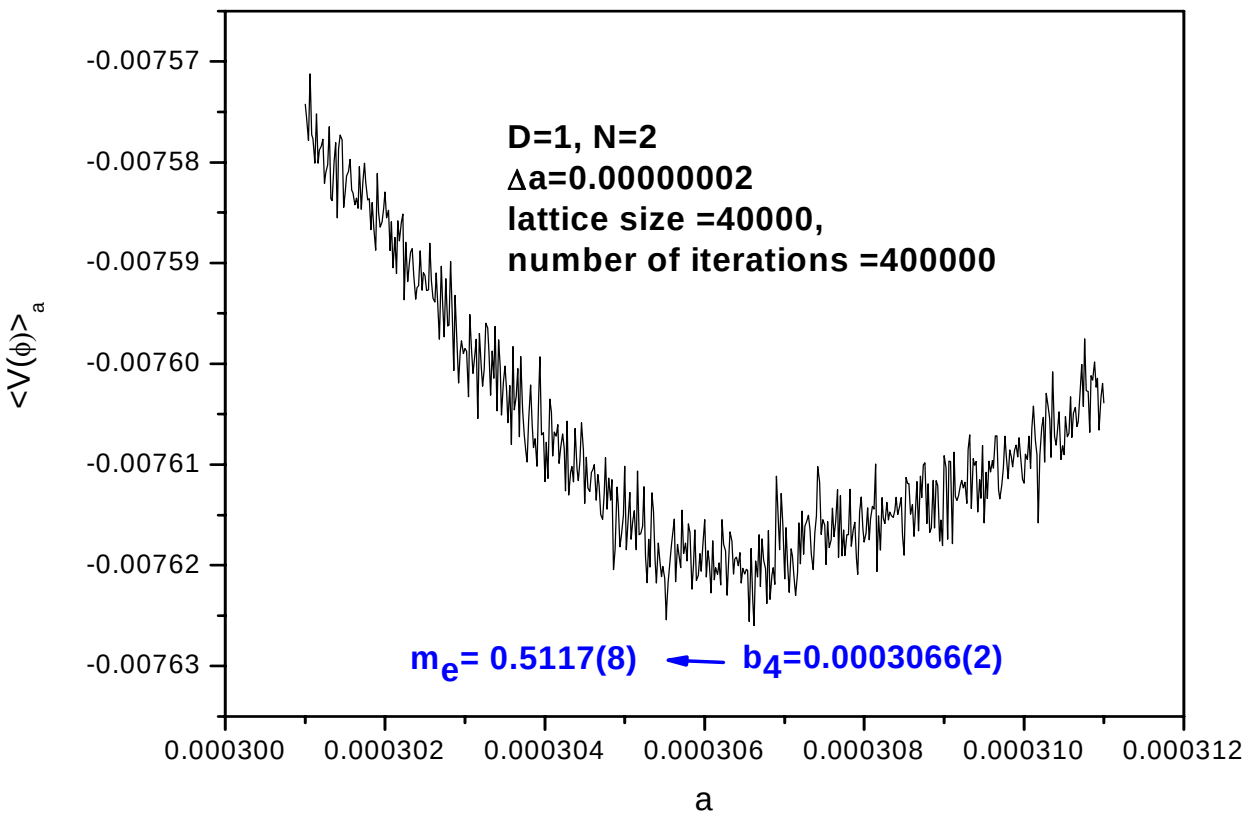

Fig (3)

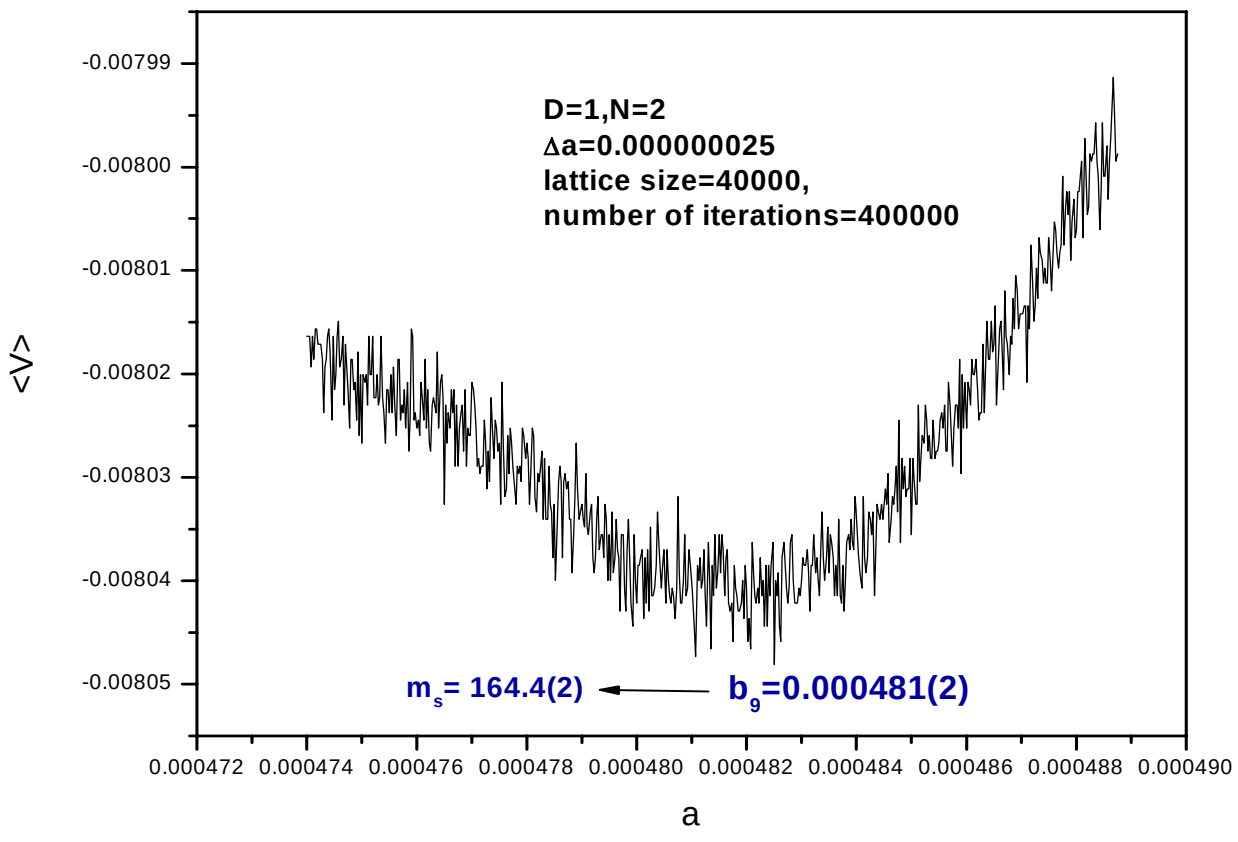

Fig (4)

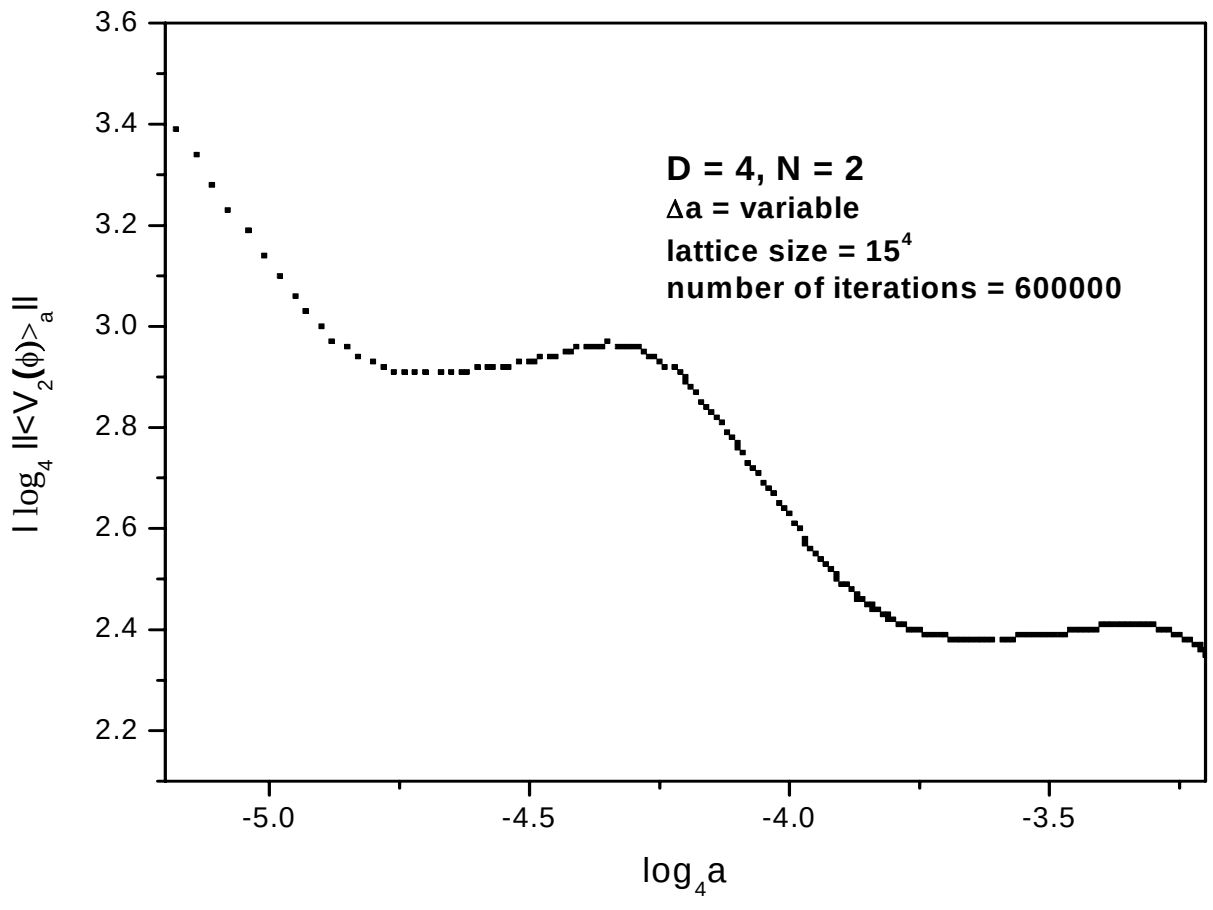

Fig (5)

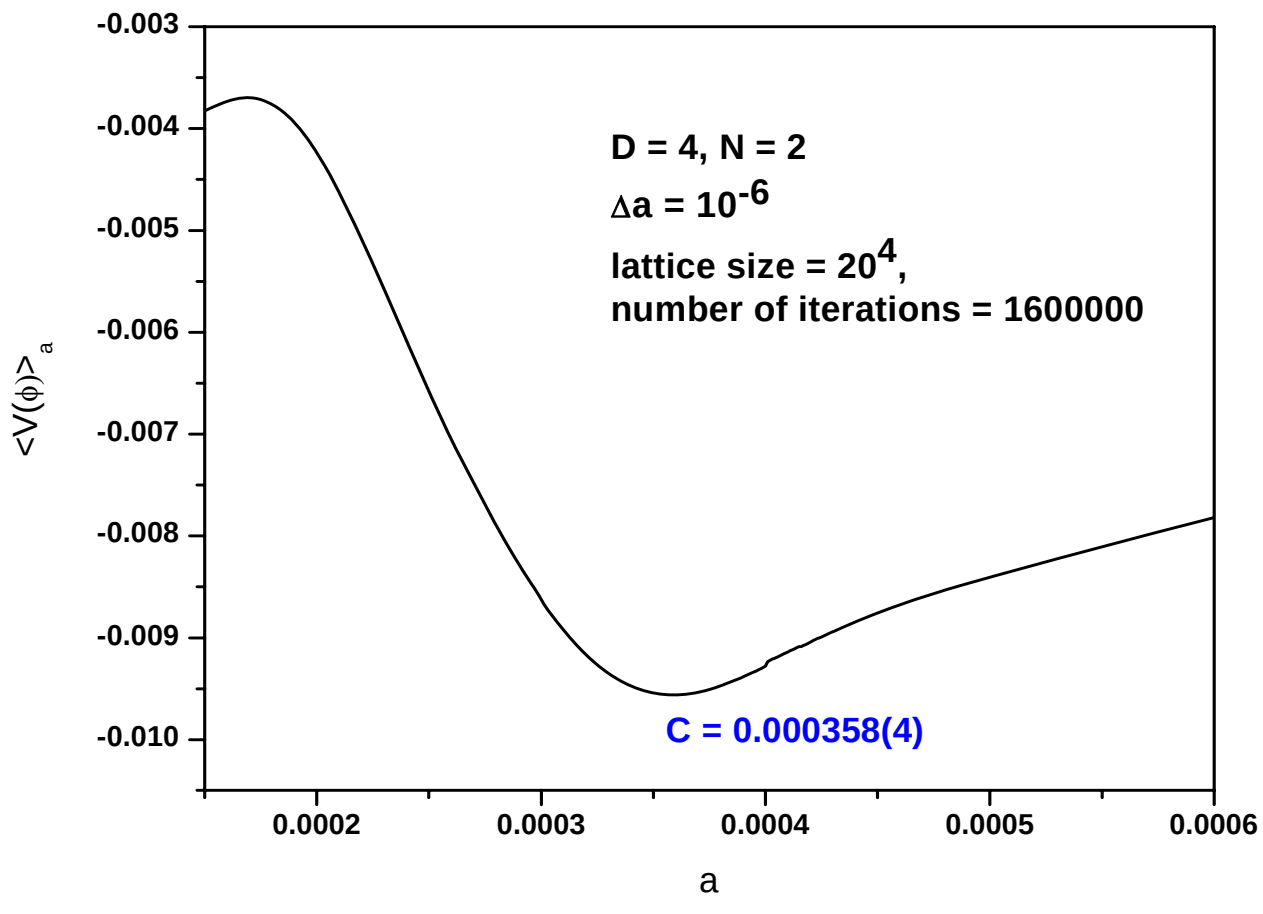

Fig (6)

## Figure Captions

Fig. 1 $\left|\log_4\left|\langle V_2(\varphi)\rangle_a\right|\right|$ versus $\log_4 a$ for the one-dimensional chaotic CML described by eq. (1) with the parameters N=2, s=1 and b=0 in the scaling region.

Fig. 2 One period of the self energy given by eq. (2) of the one-dimensional chaotic CML described by eq. (1) in the scaling region.

Fig. 3 Magnification of the local minimum labeled $b_4$ in Fig. 2.

Fig. 4 Magnification of the local minimum labeled $b_9$ in Fig. 2.

Fig. 5 $\left|\log_4\left|\langle V_2(\varphi)\rangle_a\right|\right|$ versus $\log_4 a$ for the four dimensional chaotic CML described by eq. (3) in the scaling region.

Fig. 6 One period of the self energy given by eq. (4) for the four-dimensional chaotic CML described by eq. (3) in the scaling region.